\documentclass[aps,pre,twocolumn,superscriptaddress,floats,floatfix,showpacs]{revtex4-1}
\usepackage{amssymb,amsmath}
\usepackage{float}
\usepackage{mathrsfs}
\usepackage{graphics}
\usepackage{epstopdf}
\usepackage{color}
\usepackage{subfigure}
\usepackage{epsfig}
\usepackage{rotating}
\usepackage{anyfontsize}
\usepackage{verbatim}
\usepackage{standalone}
\usepackage{hyperref}

\usepackage{xr}
\begin{document}

\title{Deep-learning-assisted detection and termination of spiral- and broken-spiral 
waves in mathematical models for cardiac tissue}
\author{Mahesh Kumar Mulimani}
\email{maheshk@iisc.ac.in ; contributed to all apsects of this study}
\affiliation{Centre for Condensed Matter Theory, Department of Physics, Indian Institute of Science, Bangalore 560012, India.}
\author{Jaya Kumar Alageshan}
\email{jayaka@iisc.ac.in ; contributed to all apsects of this study}
\affiliation{Centre for Condensed Matter Theory, Department of Physics, Indian Institute of Science, Bangalore 560012, India.}
\author{Rahul Pandit}
\email{rahul@iisc.ac.in}
\altaffiliation[\\]{also at Jawaharlal Nehru Centre For
Advanced Scientific Research, Jakkur, Bangalore, India}
\affiliation{Centre for Condensed Matter Theory, Department of Physics, Indian Institute of Science, Bangalore 560012, India.}

\pacs{87.19.Xx, 87.15.Aa }
 
\begin{abstract}

Unbroken and broken spiral waves, in partial-differential-equation (PDE) models for cardiac tissue, are the
mathematical analogs of life-threatening cardiac arrhythmias, namely,
ventricular tachycardia (VT) and ventricular-fibrillation (VF). We
develop a (a) deep-learning method for the detection of unbroken and
broken spiral waves and (b) the elimination of such waves, e.g., by the
application of low-amplitude control currents in the cardiac-tissue
context. Our method is based on a convolutional neural network (CNN)
that we train to distinguish between patterns with spiral waves
$\mathscr{S}$ and without spiral waves $\mathscr{NS}$. We obtain these
patterns by carrying out extensive direct numerical simulations (DNSs)
of PDE models for cardiac tissue in
which the transmembrane potential $V$, when portrayed via pseudocolor
plots, displays patterns of electrical activation of types
$\mathscr{S}$ and $\mathscr{NS}$. We then utilise our trained CNN to
obtain, for a given pseudocolor image of $V$, a heat map that has high
intensity in the regions where this image shows the cores of spiral
waves. Given this heat map, we show how to apply low-amplitude Gaussian
current pulses to eliminate spiral waves efficiently. Our \textit{in
silico} results are of direct relevance to the detection and
elimination of these arrhythmias because our elimination of unbroken or
broken spiral waves is the mathematical analog of low-amplitude
defibrillation.

\end{abstract}

\maketitle

The normal pumping of blood by mammalian hearts is initiated by electrical
waves of excitation that propagate through cardiac tissue and induce cardiac
contractions. The abnormal propagation of such waves can lead to cardiac
arrhythmias, like ventricular tachycardia (VT) and ventricular fibrillation
(VF), which cause sudden cardiac death (SCD) that is among the leading causes
of death in the industrialised
world~\cite{mehra2007global,Majumder2011,clayton2011models} (see, e.g.,
Refs.~\cite{honnekeri2014sudden,zheng2001sudden} for SCD data from India and
the USA). The principal cause of VT and VF are spiral or scroll waves of
electrical activation in cardiac tissue; unbroken (broken) spiral or scroll
waves are associated with VT (VF). Such waves have been studied \textit{in
vivo}~\cite{berul1996vivo,chinushi2003mechanism,gelzer2008dynamic} in mammalian
hearts, \textit{in
vitro}~\cite{davidenko1990sustained,ikeda1996mechanism,valderrabano2000obstacle,lim2006spiral}
in cultures of cardiac myocytes, and \textit{in
silico}~\cite{shajahan2007spiral,shajahan2009spiral,nayak2014spiral}, in
mathematical models for cardiac tissue. The efficient elimination of such
spiral or scroll waves and the subsequent restoration of the normal rhythm of a
mammalian heart, is a difficult problem; this can be attempted by
pharmacological means~\cite{moreno2013ranolazine} or by electrical means called
\textit{defibrillation}~\cite{sinha2001defibrillation}. Defibrillation by the
application of low-amplitude current pulses is the grand-challenge
here~\cite{luther2011low}. Two important steps are required for such
defibrillation: (a) An efficient detection of spiral waves or their broken-wave
forms; (b) the elimination of such waves by low-amplitude electrical
pulses~\cite{sinha2001defibrillation,shajahan2007spiral,shajahan2009spiral,nayak2014spiral}
or through optogenetic methods~\cite{bingen2014light}.

We develop a deep-learning method, based on a convolutional neural network
(CNN), that helps us to accomplish task (a). We then develop the mathematical
analog of a defibrillation scheme for the efficient elimination of well-formed
spiral and broken spiral waves in two dimensions (2D).  We note, in passing,
that  electrical waves in cardiac tissue belong to a large class of nonlinear
waves in excitable media, e.g., calcium-ion waves in Xenopus
oocytes~\cite{lechleiter1991spiral}, waves in chemical reactions of the
Belousov-Zhabotinsky type~\cite{winfree1972spiral}, waves that occur during the
oxidation of carbon monoxide on the surface of
platinum~\cite{falcke1992traveling,imbihl1995oscillatory,pande1999spatiotemporal},
excitable-wave patterns in a recent semiconductor-laser
experiment~\cite{marino2019excitable}, and waves in dictyostelium discoideum
that are associated with the cyclic-AMP
signalling~\cite{tyson1989cyclic,rietdorf1996analysis}; our step (a) can be
applied, \textit{mutatis mutandis}, for the detection of spiral waves in such
systems.

\begin{figure}[!ht]
\includegraphics[scale=0.4]{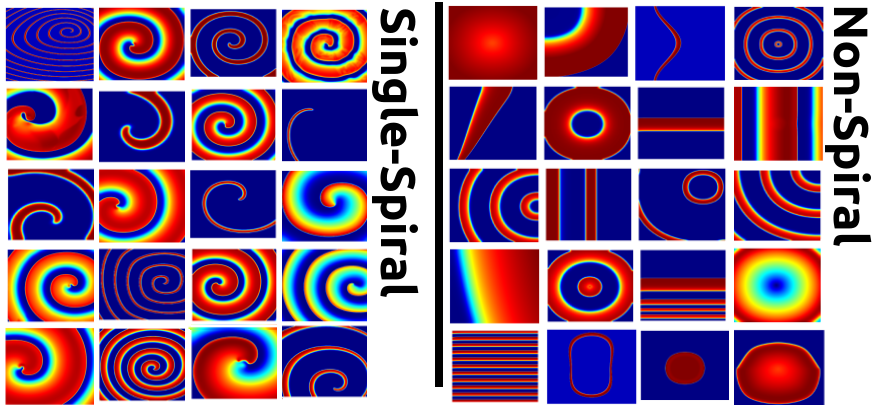}
		\put(-148,125) {\bf(a)}
		\put(-17,125) {\bf(b)}	
\caption{Collages of illustrative pseudocolor plots of $V$ with (a) 
single-spiral ($\mathscr{S}$) images and (b) no spiral ($\mathscr{NS}$) images, 
which we use for training our CNN. In (a), the arm-width of the spiral and
its rotation frequency depends on the model [Eqs.~(\ref{eqn:twovar}) or (\ref{eqn:pde})]
and the parameters therein; the spiral-wave-initiation scheme controls the chirality of 
the spiral and the position of its core. In (b), the $\mathscr{NS}$ class 
includes plane waves and target waves.}
		
\label{fig:collage}
\end{figure}

Specifically, we \textit{train} our CNN \textit{to classify}, into the following two 
sets, patterns of electrical-wave activation, which we obtain from 
\textit{in silico} studies of different mathematical models for cardiac
tissue~\cite{barkley1991model,aliev1996simple,luo1991model,ten2006alternans,
o2011simulation}: (a) spiral waves ($\mathscr{S}$); and
(b) no spiral waves ($\mathscr{NS}$) (Fig.~\ref{fig:collage}).
Next, we use our trained CNN to detect spiral-wave patterns, with both unbroken 
and broken spirals. We then use the outputs from our CNN to construct a 
\textit{heat map} that has high intensity in the regions with spiral cores. 
We demonstrate how to eliminate the broken or unbroken
spiral waves by applying low-amplitude current stimuli at those positions 
at which the heat map has high intensity; this is the mathematical analog 
of \textit{defibrillation}~\cite{sinha2001defibrillation}.

\begin{figure}[!ht]
\includegraphics[scale=0.35]{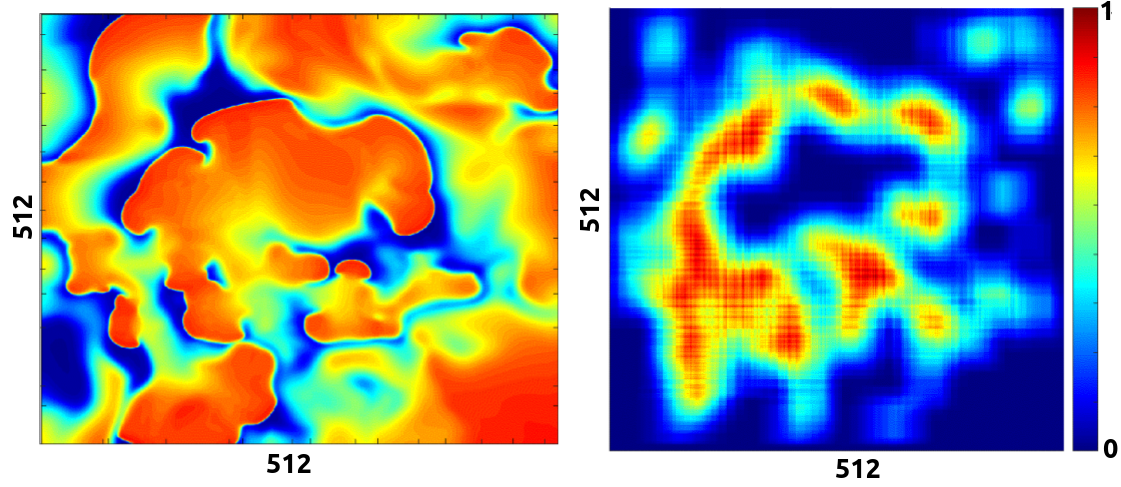}
	\caption{Pseudocolor plots of (left panel) $V$, showing broken spiral waves, 
	and (right panel) the heat map $\mathcal{H}$ for the image in the left panel
	(see text).}
\label{heatmap}
\end{figure}

Mathematical models for cardiac tissue use nonlinear partial differential
equations (PDEs) of the reaction-diffusion type given below:
\begin{eqnarray}
   \frac{\partial V}{\partial t} = D_0 \nabla^{2}V + f(V,g);  \;\; 
   \label{eqn:twovar}
   \frac{\partial g}{\partial t} = \epsilon(V,g)  h(V,g); \\
  \frac{\partial{V}}{\partial{t}} = D_0 {\nabla^2}V - \frac{I_{ion}}{C_{m}}; \;\;\;\;\;\
	I_{ion} = \sum_{i} I_i. \;\;\;\;\;\;\;\;\;\;\;\
   \label{eqn:pde}
\end{eqnarray}
We use two classes of models, namely, (a) two-variable models
[Eq.(\ref{eqn:twovar})] and (b) biologically realistic models with ion
channels, ion pumps, and ion exchangers [Eq.(\ref{eqn:pde})]. The type-(a)
models that we work with are (i) the Barkley model~\cite{barkley1991model} and
(ii) the Aliev-Panfilov model~\cite{aliev1996simple}, in which $V$, $g$, and
$D_0$ are, respectively, the transmembrane potential, the effective ionic gate
(a slow variable), and the diffusion constant; $f$ and $h$ are nonlinear
functions of $V$ and $g$.  We employ the following type-(b) models: (i) the
Luo-Rudy phase-I (LR-I) guinea-pig-ventricular model~\cite{luo1991model}; (ii)
the TP06 human-ventricular model~\cite{ten2006alternans}; and (iii) the
O'Hara-Rudy (ORd) human-ventricular model~\cite{o2011simulation}, where $V$,
$I_{i}$, and $C_m$ are the transmembrane potential, the ionic current for
ion-channel $i$, and the membrane capacitance, respectively (see the
Supplemental Material~\cite{Supmat_Final_2019} for the forms of $f, \, h$, and $I_i$ in
these models).  We use the forward-Euler method (step size $\Delta t$) for time
marching, and a finite-difference scheme in space (step size $\Delta x$), with
a 5-point stencil for the Laplacian, and no-flux boundary conditions to obtain
numerical solutions of Eqs.~(\ref{eqn:twovar}) and (\ref{eqn:pde}), in
two-dimensional (2D) square domains with $N\times N$ grid points, with $128 \le N \le 1024$
; in most 
of our simulations we use $N=512$. We choose
$\Delta t$ and $\Delta x$ such that the von-Neumann stability criterion is
satisfied~\cite{press2007numerical}.  We focus on electrical activity in
cardiac tissue, so, from our numerical solutions of these PDEs, we extract the spatiotemporal
evolution of $V$, which yields several patterns like \textit{spiral waves}
(Fig.\ref{fig:collage}(a)), \textit{target waves} (top right corner in
Fig.\ref{fig:collage}(b)), \textit{\it plane waves} (bottom left corner in
Fig.\ref{fig:collage}(b)), and states with \textit{\it spiral break-up}
(Fig.\ref{heatmap}(a)). 

In Fig.~\ref{fig:collage} we show representative pseudocolor images of $V$ that
we obtain from our numerical solutions of Eqs.~(\ref{eqn:twovar}) and
(\ref{eqn:pde}) over a wide range of parameters in the cardiac-tissue models
that we have listed above. We use 22,000 such images to train, and then test,
our CNN. We create additional images by performing geometrical operations on
the primary pseudocolor images $V$, e.g., inequivalent reflections about the
horizontal and vertical axes, so that our dataset of images is not biased in
favor of any particular orientation; this improves the training performance of our
CNN. We train our CNN with $70\%$ of the total number of images; and we save
the remaining $30\%$ of the images for the validation of our CNN model.

\begin{figure*}[!ht]
\includegraphics[scale=0.4]{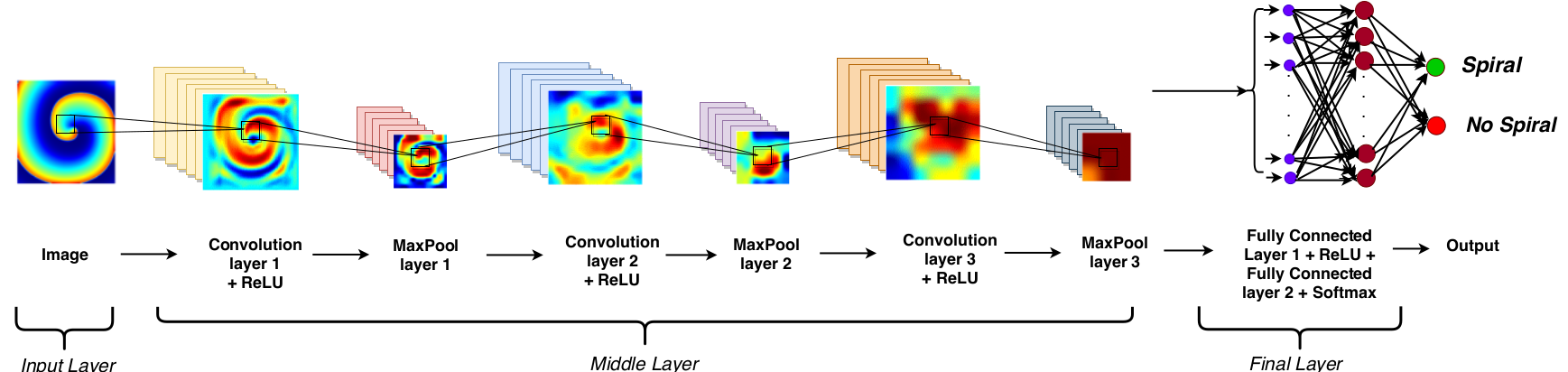}
\caption{A schematic diagram of our CNN showing the Input, Middle, and Final layers; 
the details of each one of these layer are given in the Supplemental 
Material~\cite{Supmat_Final_2019}.}
\label{fig:cnn}
\end{figure*}

Our solutions of Eqs.~(\ref{eqn:twovar}) or (\ref{eqn:pde}) yield $V$ at $N^2$
grid points. We first define the normalised transmembrane potential
$\tilde{V}=(V-V_{min})/(V_{max}-V_{min})$; $V_{max}$ and $V_{min}$ are,
respectively, the maximal and minimal values of $V$, so $0\le \tilde{V} \le 1$.
We then reduce the large number of grid points by specifying $\tilde{V}$  on
$32 \times 32$ points by using the \textit{resize} function in MATLAB R2018b.
We use the Deep Learning Toolbox in MATLAB R2018b to develop our CNN, which we
depict schematically in Fig.~\ref{fig:cnn}. It has three main layers: (1)
\textit{Input}; (2) \textit{Middle}; and (3) \textit{Final}.  The Middle layer
contains three sets of \textit{Convolution}, \textit{Rectified Linear Unit}
(ReLU), and \textit{MaxPool} sub-layers.  The Final layer contains two fully
connected \textit{Artificial Neural Networks} (ANNs). We give a brief
description of the implementation of our CNN in the Supplemental
Material~\cite{Supmat_Final_2019}.

We begin the training by feeding the the image of $\tilde{V}$ to our CNN. If
the CNN output predicts the class of the input image incorrectly, then we use a
proxy cost function to rectify this error iteratively (until the CNN yields the
correct output class). Specifically, we achieve this for our CNN by minimizing
the cross-entropy cost function
\begin{equation}
\small{\mathcal{C} = -\sum_{\ell=1}^{M} \sum_{q=0}^{\eta-1} \frac{\large{[} O_{\ell,q} 
	\ln(\tilde{O}_{\ell,q}) + (1-O_{\ell,q}) \ln(1-\tilde{O}_{\ell,q})\large{]}}{M}},
\label{eq:crossentropy}
\end{equation}
by using the stochastic-gradient-descent method with a learning rate
$\alpha=0.001$ (see, e.g., Chapter 2 of Ref~\cite{nielsen2015neural}); here,
$\tilde{O}_{\ell,q}$ are the CNN outputs ($\tilde{O}_{\ell,q} \in (0,1)$) and
$O_{\ell,q}$ are the real outputs, for the input image $\ell$, and $M$ is the
\textit{mini-batch} size (the total number of images is divided into subsets,
called mini-batches, with $M$ images each); we use $M=128$. For the class
$\mathscr{S}$, $O_{\ell,0} =1$ and $O_{\ell,1} =0$; and for $\mathscr{NS}$,
$O_{\ell,0} =0$ and $O_{\ell,1} =1$.

Even though we train our CNN with single-spiral-wave patterns, it manages to
identify patterns with broken spiral waves as belonging to the class
$\mathscr{S}$: We have checked that this CNN classifies 10,000
broken-spiral-wave patterns (see the pseudocolor plot of $V$ in
Fig.\ref{heatmap} (a)) as $\mathscr{S}$, with an accuracy of $99.6\%$.  This is
especially useful when we carry out the mathematical analog of defibrillation,
i.e., the elimination of all spirals, unbroken or broken. We can, indeed,
utilise our trained CNN to examine pseudocolor plots of $V$, during our
numerical simulation of a mathematical model for cardiac tissue; the moment
this CNN detects a pattern of type $\mathscr{S}$, we can eliminate it by the
application of suitable currents on a coarse, control
mesh~\cite{sinha2001defibrillation} or by using an optogenetics-based
control~\cite{bingen2014light} method (we give a brief description of these
control methods (see Fig. S3) in the Supplemental Material~\cite{Supmat_Final_2019}).
Here, we discuss a new scheme for eliminating both broken and unbroken spiral
waves; this relies on developing a \textit{heat map}, from a pseudocolor plot
of $V$, for those images that are identified by our CNN to lie in the class
$\mathscr{S}$.  This heat map (Fig.~\ref{heatmap} (b)) is 
\begin{equation}
	\mathcal{H}(i,j) = \frac{N_p}{N} \sum_{r=1}^{N/Np} {\rm CNN}\left(\chi^r_{i,j}
	        \left\{ V \right\} \right),  
\label{eq:hp}
\end{equation} 

$\forall i, j\in\{1,2, \ldots, N\}$; the arguments of the matrix-resizing
function $\chi^r_{i,j}$ (a standard function in Matlab) are the $32 r \times 32
r$ values of $V$ in a square of side $32 r$ centred at the point $(i,j)$, with
$1 \leq i, j \leq N$ and $1 \leq r \leq (N/N_p)$; for the images we employ,
$N=512$ and $N_p=32$. We use the resized $\chi^r_{i,j}$, an image with $N_p
\times N_p$ pixels, as an input into our CNN (Fig.~\ref{fig:cnn}) and its
output, $0$ (for $\mathscr{NS}$) or $1$ (for $\mathscr{S}$), is summed over $r$
to obtain $\mathcal{H}(i,j)$ for a given input pseudocolor plot of $V$.
Clearly, $\mathcal{H}(i,j)\in[0,1]$; and it is large if there is a spiral core
near the point $(i,j)$.  In the left and right panels of Fig.~\ref{heatmap} we
depict, respectively, a pseudocolor plot of $V$, with a broken-spiral-wave
pattern, and the corresponding heat map.  We now show how to use such a heat
map to develop a new control scheme for the elimination of both broken and
unbroken spiral waves.

\begin{figure*}[!ht]
\includegraphics[scale=0.45]{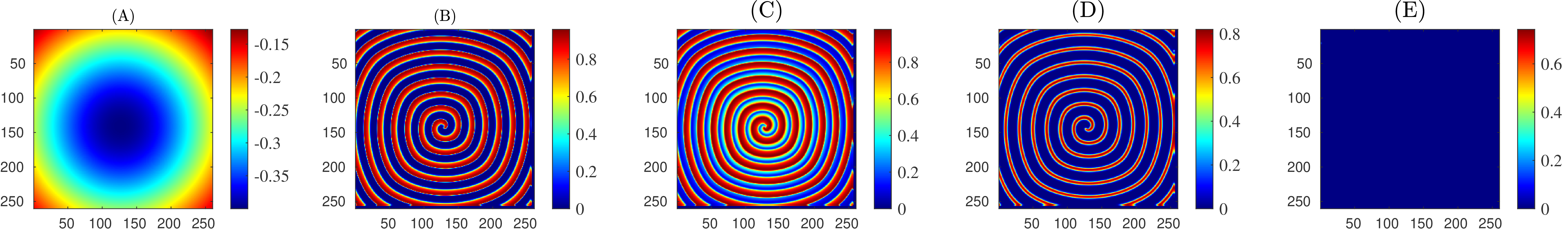}
\caption{Pseudocolor plots which shows, (A) the defibrillation current as $\mathcal{P_H}$ (single 2D Gaussian centered at the spiral core), whose $I_{def} = 0.5$ pA/pF, $\sigma = 0.375 N \Delta x$ (cm) and $N=256 $. (B)-(F) $V$ of the Aliev-Panfilov model (Ref~\cite{aliev1996simple}). (B) the spiral wave before application of the defibrillation current and (C) shows the time of application and (D)-(F) After the defibrillation current is applied. We observe that spiral wave is eliminated after the application of the defibrillaion current of 2D Gaussian profile.}
\label{gaussspiral}
\end{figure*}

Spiral-wave excitations emanate from the spiral core, so we might expect 
that the elimination of this core could lead to the removal of 
the spiral wave. However, when we apply a current pulse
on a disk centred at the spiral core, we find that it leads to the formation 
of multiple spiral cores along the boundary of the disk. We prevent the formation of
such multiple spiral cores as follows:
We first show, for a single spiral wave in the Aliev-Panfilov model, that, by applying 
a current pulse with a spatial profile that is a symmetrical, two-dimensional (2D) 
Gaussian (centred at the spiral core, with equal widths in both $x$ and $y$ directions 
$\sigma_x = \sigma_y = \sigma $, and with a peak intensity of $ I_{def} $), 
we can remove the core and the wave \textit{without forming multiple spiral cores} 
(Fig.~\ref{gaussspiral}). Henceforth, we refer to such 
a current profile as a Gaussian current pulse with width $\sigma$.

We now consider a pattern with multiple spiral waves (e.g., the pseudocolor
plot of $V$ in the left panel of Fig.~\ref{heatmap}), whose spatial extent is
much smaller than the large spiral wave in Fig.~\ref{gaussspiral}.  We consider
a square lattice of points, labelled by $(i',j')$, with $0 \leq i', j' \leq
N_G$, i.e., the side of the unit cell $a = N_G \Delta x$ (cm). On each of these
points we impose a Gaussian current pulse $G(i',j')$ of width $\sigma$ and
amplitude $I_{def}$; the total, normalised contribution of these pulses, at the
point  $(i,j)$, with $0 \leq i, j \leq (N-1)$, in the original image, is
\begin{eqnarray}
	\tilde{\mathcal{G}}(i,j) &=& 
	\sum_{i',j'=0}^{N_G} G\left(i-i'\frac{N}{N_G}, j-j'\frac{N}{N_G}\right); \nonumber \\
	\mathcal{G}(i,j) &=& \tilde{\mathcal{G}}(i,j)/[\tilde{\mathcal{G}}_{max}]; 
\label{eq:sumGauss}
\end{eqnarray}
here, $\tilde{\mathcal{G}}_{max} = max_{(i,j)}[\tilde{\mathcal{G}}(i,j)]$.  The
final current pulse that we apply, for a time $t_{def}$ ms  at the point
$(i,j)$, is given by the Hadamard product 
\begin{equation}
\mathcal{P_H}(i,j) = I_{def} \ \mathcal{H}(i,j)\mathcal{G}(i,j),
\label{eq:Hadamard}
\end{equation}
where $I_{def}$ sets the scale of the current that is applied. 
Multiple spiral waves are eliminated by the application of  
$\mathcal{P_H}(i,j)$ (henceforth, Gaussian-control scheme) 
as we demonstrate below.

\begin{figure*}[!ht]
\includegraphics[scale=0.4]{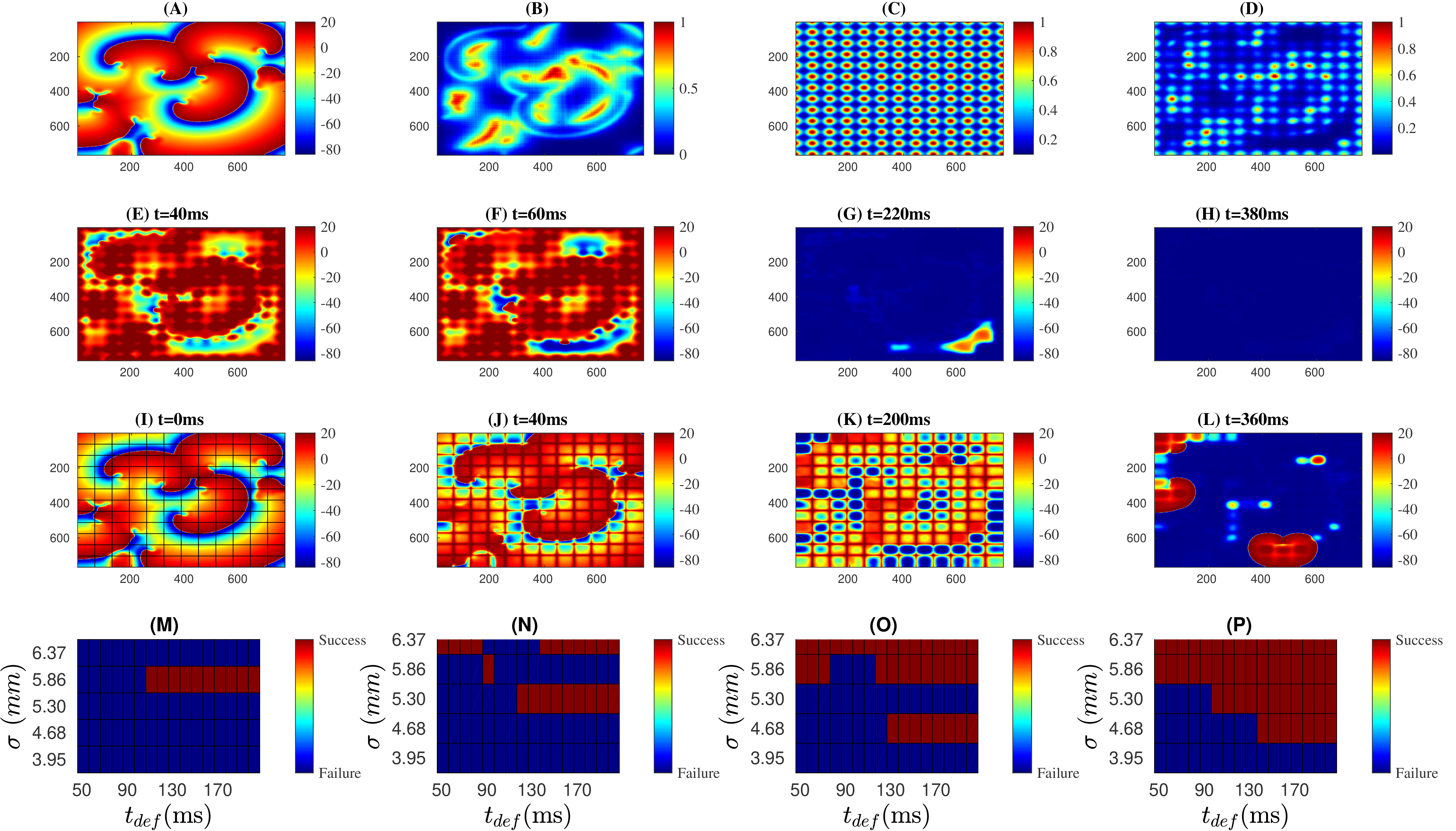}
\hspace{2.5cm}

	\caption{Pseudocolor plots of (A) $V$, for an illustrative, broken
	spiral wave from our simulation of the the TP06 model, (B) the
	corresponding heat map $\mathcal{H}$ (Eq.~\ref{eq:hp}), (C) the summed
	and normalised 2D Gaussians ($\mathcal{G}(i,j)$ in
	Eq.~\ref{eq:sumGauss}), and (D) the Hadamard product
	$\mathcal{P_H}(i,j)$ (with $I_{def} = 1$ pA/pF in
	Eq.~\ref{eq:Hadamard}). (E)-(H): Pseudocolor plots of $V$, at different
	representative times, showing the elimination of the broken spiral
	waves (as in (A) at $t=0$) after the application of
	$\mathcal{P_H}(i,j)$; $\mathcal{P_H}(i,j)$ is turned off at $t =
	t_{def} = 120$ ms (for the complete spatiotemporal evolution see the
	video V1 in the Supplemental Material~\cite{Supmat_Final_2019}). (I)-(L): the
	analogs of (E)-(H) for the current-mesh control scheme of
	Refs.~\cite{sinha2001defibrillation} (see the video V2 in Supplemental
	Material~\cite{Supmat_Final_2019}). Phase diagrams in the $(t_{def}, \sigma)$
	plane showing parameter regions in which our Gaussian control scheme
	[(E)-(F)] succeeds (red) and does not succeed (blue) in controlling
	broken spiral waves for (M) $I_{def} = 5$ pA/pF, (N) $I_{def} =
	10$ pA/pF, (O) $I_{def} = 15$ pA/pF and (P) $I_{def} = 20$ pA/pF.}

\label{sp_hp_gaus}

\end{figure*}

In Figs.~\ref{sp_hp_gaus} (A) and (B), we show illustrative pseudocolor plots
of, respectively, $V$ and its heat map $\mathcal{H}$ (Eq.(~\ref{eq:hp})), for an
image with a broken spiral wave, in the TP06 model. Figures~\ref{sp_hp_gaus} (C)
and (D) depict, respectively, pseudocolor plots of the summed and normalised 2D
Gaussians ($\mathcal{G}(i,j)$ in Eq.(~\ref{eq:sumGauss})) and the Hadamard
product $\mathcal{P_H}(i,j)$ (with $I_{def} = 1$ pA/pF in
Eq.(~\ref{eq:Hadamard})). Our Gaussian-control scheme is illustrated by the
pseudocolor plots of $V$ [Figs.~\ref{sp_hp_gaus} (E)-(H)]; these show the
spatiotemporal evolution of $V$ after the application of $\mathcal{P_H}(i,j)$,
which is turned off at $t = t_{def}$ (for the complete spatiotemporal
evolution see the video V1 in the Supplemental Material~\cite{Supmat_Final_2019}).  In
Figs.~\ref{sp_hp_gaus} (I)-(L) we show the counterparts of
Figs.~\ref{sp_hp_gaus} (E)-(H) for the control scheme in which a current pulse
is applied on a square mesh to eliminate broken spiral waves (see
Refs.~\cite{sinha2001defibrillation,shajahan2007spiral,shajahan2009spiral,nayak2014spiral},
the Supplemental Material~\cite{Supmat_Final_2019}, and the video V2);
we refer to this as the mesh-control scheme.

\begin{table}[H]
\begin{center}
\begin{tabular}{|l|l|l|l|l|l|}
\hline
Control & \ $N_{G}$ \ & \ \ $a$ \  \ & \ \ \ $\sigma$ \ \ \ & \ \ $I_{def} $ \ \ & \ \ $t_{def}$ \ \\
scheme & \  \ & (cm) & \ (cm) \ & (pA/pF) & \ (ms) \ \\
\hline
\hline
 GC1 & \ 64 \ & \ 1.6 \ & \ 0.37 $a$ \ & \ \ \ \ 5 & 120 \\
 GC2 & \ 96 \ & \ 2.4 \ & \ 0.37 $a$ \ & \ \ \ \ 5 & 120 \\
 MC & \ 64 \ & \ \ \ - \ & \ \ \ - \ & \ \ \ \ 15 &  120 \\
\hline
\end{tabular}
\end{center}
	\caption{The parameter values that we use for our Gaussian-control (GC1 and GC2)
and Mesh-control (MC) schemes in our illustrative simulations.}
\label{Table:table1}
\end{table}

The efficacy of our Gausssian-control scheme depends on the parameters $a, \,
\sigma, \, I_{def}$, and $t_{def}$. We list these parameters
(Table~\ref{Table:table1}) for two illustrative Gaussian-control runs, GC1 and
GC2, and one run, MC, in which we use a mesh-control scheme, for the TP06
model. By comparing the results of such runs we find that, for large values of
$a$, our Gaussian-control scheme is not successful in removing spiral waves;
e.g., in the TP06 model, broken spiral waves are suppressed for the value of
$a$ that we use in run GC1, but not for the value of $a$ in run GC2
(Table~\ref{Table:table1}).  For the parameters in run GC1,
Figs.~\ref{sp_hp_gaus} (M)-(P) show phase diagrams, in the $(t_{def}, \sigma)$
plane and for representative values of $I_{def}$, with parameter regions in
which our Gaussian-control scheme succeeds (red) and does not succeed (blue) in
controlling broken spiral waves: This Gaussian-control scheme also eliminates
broken and unbroken spiral waves in all the other cardiac-tissue models that we
have studied (see Figs. S5 and S6 in the Supplemental Material~\cite{Supmat_Final_2019}).

By comparing the pseudocolor plots in rows two and three of
Fig.~\ref{sp_hp_gaus}, we can contrast the effectivness of our Gaussian-control
scheme with that of the mesh-control scheme of
Refs~\cite{sinha2001defibrillation}. We find, in particular, that our Gaussian-control scheme eliminates broken spiral waves with $I_{def} = 5$ pA/pF; by
contrast, the mesh scheme requires $I_{def} = 15$ pA/pF for such
elimination. Thus, the Gaussian-control scheme leads to the elimination of
spiral waves with lower local currents than the mesh-control scheme, with all
other parameters held fixed.

We have checked that our CNN can be used to detect scroll waves in
three-dimensional(3D) simulation domains (see Fig. S4 of the Supplimentary
Material Ref~\cite{Supmat_Final_2019}). The elimination of such 3D
scroll waves by the application of currents on a 2D surface of a 3D domain
remains a significant challenge
Ref.~\cite{sinha2001defibrillation,shajahan2007spiral,shajahan2009spiral,nayak2014spiral}.

Our deep-learning-assisted Gaussian-control method is an important step in the
detection and elimination of both broken and unbroken spiral waves.
Machine-learning techniques have been used, e.g., in
Refs.~\cite{figuera2016machine,marvsanova2017ecg,hannun2019cardiologist}, for
the effective detection of anomalies in electrocardiograms (ECGs), which can
then be eliminated by the controlled delivery of electrical signals via
automated defibrillators (see, e.g.
Refs.~\cite{figuera2016machine,singh2018classification,SitengC}). To the best
our of knowledge, no machine-learning method has been employed so far for the
detection of spiral waves in, e.g., pseudocolor plots of $V$. Our study uses
the complete spatial information in patterns of $V$ to develop an efficient
Gaussian-control scheme for the elimnation of unbroken and broken spiral waves,
which are the mathematical analogs of life-threatening VT and
VF~\cite{shajahan2007spiral,shajahan2009spiral,nayak2014spiral}.

We hope that our CNN-based detection of spiral waves and our Gaussian-control
scheme will be tested in \textit{in-vitro} experiments with cardiac myocytes,
such as those used in the studies of
Refs.~\cite{davidenko1990sustained,ikeda1996mechanism,valderrabano2000obstacle,lim2006spiral,bingen2014light,shajahan2016scanning,kudryashova2017virtual}.
Our CNN can also be used to detect spiral waves in other excitable media~\cite{lechleiter1991spiral,winfree1972spiral,falcke1992traveling,imbihl1995oscillatory,pande1999spatiotemporal,marino2019excitable,tyson1989cyclic,rietdorf1996analysis}.

\section*{ Acknowledgments}

{We thank DST, CSIR, UGC (India) for the support and the Supercomputer Education
and Research Centre (SERC, IISc) for computational resources.}

\bibliographystyle{apsrev}
\bibliography{reference}

\end{document}


\title{Supplementary Material: Deep-learning-assisted detection and termination of spiral and broken spiral waves
in mathematical models for cardiac tissue}
\author{Mahesh Kumar Mulimani}
\email{maheshk@iisc.ac.in}
\affiliation{Centre for Condensed Matter Theory, Department of Physics, Indian Institute of Science, Bangalore 560012, India}
\author{Jaya Kumar Alageshan}
\email{jayaka@iisc.ac.in}
\affiliation{Centre for Condensed Matter Theory, Department of Physics, Indian Institute of Science, Bangalore 560012, India}
\author{Rahul Pandit}
\email{rahul@iisc.ac.in}
\altaffiliation[\\]{also at Jawaharlal Nehru Centre For
Advanced Scientific Research, Jakkur, Bangalore, India}
\affiliation{Centre for Condensed Matter Theory, Department of Physics, Indian Institute of Science, Bangalore 560012, India}
\pacs{87.19.Xx, 87.15.Aa }

\maketitle

We provide details of (a) the models we use [Sec.~\ref{sec:models} ], (b) our
convolutional neural network (CNN) [Sec.~\ref{sec:CNN} ], and (c) some
low-amplitude defibrillation (or spiral-removal) schemes [Sec.~\ref{sec:defib} ]
in mathematical models for cardiac tissue.

\section{Models}

As we have mentioned in the main paper, mathematical models for cardiac tissue
use nonlinear partial differential equations (PDEs) of the reaction-diffusion
type given below:

\begin{eqnarray}
   \frac{\partial V}{\partial t} = D_0 \nabla^{2}V + f(V,g);  \;\; 
   \label{eqn:twovar}
   \frac{\partial g}{\partial t} = \epsilon(V,g)  h(V,g); \\
  \frac{\partial{V}}{\partial{t}} = D_0 {\nabla^2}V - \frac{I_{ion}}{C_{m}}; \;\;\;\;\;\
	I_{ion} = \sum_{i} I_i. \;\;\;\;\;\;\;\;\;\;\;\
   \label{eqn:pde}
\end{eqnarray}
We use two classes of models, namely, (a) two-variable models
[Eq.(\ref{eqn:twovar})] and (b) biologically realistic models with ion
channels, ion pumps, and ion exchangers [Eq.(\ref{eqn:pde})]. The type-(a)
models that we work with are (i) the Barkley model~\cite{barkley1991model} and
(ii) the Aliev-Panfilov model~\cite{aliev1996simple}, in which $V$, $g$, and
$D_0$ are, respectively, the transmembrane potential, the effective ionic gate
(a slow variable), and the diffusion constant; $f$ and $h$ are nonlinear
functions of $V$ and $g$.  We employ the following type-(b) models: (i) the
Luo-Rudy phase-I (LR-I) guinea-pig-ventricular model~\cite{luo1991model}; (ii)
the TP06 human-ventricular model~\cite{ten2006alternans}; and (iii) the
O'Hara-Rudy (ORd) human-ventricular model~\cite{o2011simulation}, where $V$,
$I_{i}$, and $C_m$ are the transmembrane potential, the ionic current for
ion-channel $i$, and the membrane capacitance, respectively. We give the
the forms of $f, \, h$, and $I_i$ in these models below. 
\label{sec:models}

{\it Models of Type (a)}

\begin{eqnarray}
   \frac{\partial V}{\partial t} &=& D_0 \nabla^{2}V + f(V,g); \\
   \frac{\partial g}{\partial t} &=&  h(V,g).
\end{eqnarray}

{\it (i) Barkley (B) model}
\begin{eqnarray}
   f(V,g) &=& -\frac{1}{\epsilon_0}V(1-V)(V-V_{thr}(g)) ; \nonumber\\
   h(V,g) &=& V-g ;\nonumber\\
    \epsilon_{0} = \frac{\tau_{V}}{\tau_{g}} ;\nonumber
    V_{thr}= \frac{g+b}{a} .
   \label{eqn:bm}
\end{eqnarray}

{\it (ii) Aliev-Panfilov (AP) model}
\begin{eqnarray}
   f(V,g) &=& -kV (V-V_{thr})  (V-1) -V g ; \nonumber\\
   h(V,g) &=& \epsilon(V,g) (-g-k  V  (V-V_{thr}-1)) ;  \nonumber\\
   \epsilon(V,g) &=& \epsilon_{0} + \frac{\mu_{1}  g}{\mu_{2} + V} ; 
      \;\; \epsilon_{0} =     
      \frac{\tau_{V}}{\tau_{g}} .
   \label{eqn:apm}
\end{eqnarray}

{\it Models of Type (b)} 
\begin{eqnarray}
  \frac{\partial{V}}{\partial{t}} = D_0 {\nabla^2}V - \frac{I_{ion}}{C_{m}}; \;\;\;\;\;\
	I_{ion} = \sum_{i} I_i \;\;\;\;\;\;\;\;\;\;\;\
   \label{eqn:pde}
\end{eqnarray}

We list the currents $I_i$ for all the realistic models we use in
Tables~\ref{sec:Table1}, \ref{sec:Table2}, and \ref{sec:Table3}. The equations for 
these currents can be found in Refs.~\cite{luo1991model,ten2006alternans,o2011simulation}.

\begin{table}[!ht]
	{\it (iii) Luo-Rudy phase-I (LR-I) model}
\begin{center}
    \begin{tabular}{ | l |  p{7.5cm} |}
    \hline 
    $I_{Na}$ & fast inward sodium current  \\ \hline 
    $I_{si}$ & slow inward calcium current   \\ \hline
    $I_{K}$ & Time dependent outward potassium current   \\ \hline
    $I_{K1}$ & Time independent outward potassium current    \\ \hline
    $I_{Kp}$ & Plateau outward potassium current   \\ \hline 
    $I_{b}$ & background current    \\ \hline
    
    \end{tabular}

 \hspace{7cm}\caption{\textbf{The various ionic currents in the LR-I model; the symbols 
used for these currents follow Ref.~\cite{luo1991model}. }}
\label{sec:Table1}
\end{center}
\end{table}     

\begin{table}[!ht]
	{\it (iv) ten-Tusscher and Panfilov (TP06) model}
\begin{center}
    \begin{tabular}{ | l |  p{7.5cm} |}
    \hline 
    $I_{Na}$ & fast inward sodium current  \\ \hline 
    $I_{CaL}$ & L-type inward calcium current   \\ \hline
    $I_{to}$ & Transient outward current   \\ \hline
    $I_{Ks}$ & Slow delayed rectifier outward potassium current    \\ \hline
    $I_{Kr}$ & Rapid delayed rectifier outward potassium current   \\ \hline 
    $I_{K1}$ & Inward rectifier potassium current   \\ \hline
    $I_{NaCa}$ & $Na^+ / Ca^{++}$ exchange current  \\ \hline
    $I_{NaK}$ & $Na^+ /K^+$ pump current    \\ \hline
    $I_{pCa}$ & plateau calcium current  \\ \hline 
    $I_{pK}$ & plateau potassium current   \\ \hline
    $I_{bNa}$ & background inward sodium current   \\ \hline
    $I_{bCa}$ & background inward calcium current    \\ \hline
    
    \end{tabular}

 \hspace{7cm}\caption{\textbf{The various ionic currents in the TP06 model; the symbols 
used for these currents follow Ref.~\cite{ten2006alternans}. }}
\label{sec:Table2}
\end{center}
\end{table}

\begin{table}[!ht]
	{\it (v) O'Hara and Rudy (ORd) model} 
\begin{center}
    \begin{tabular}{ | l |  p{7.5cm} |}
    \hline 
    $I_{Na}$ & inward sodium current  \\ \hline 
    $I_{CaL}$ & L-type inward calcium current   \\ \hline
    $I_{to}$ & Transient outward potassium current   \\ \hline
    $I_{Ks}$ & Slow delayed rectifier outward potassium current    \\ \hline
    $I_{Kr}$ & Rapid delayed rectifier outward potassium current   \\ \hline 
    $I_{K1}$ & Inward rectifier outward potassium current   \\ \hline
    $I_{NaCa}$ & $Na^+ / Ca^{++}$ exchange  current  \\ \hline
    $I_{CaNa}$ & CaMK activated sodium current  \\ \hline
    $I_{CaK}$ & CaMK activated potassium current    \\ \hline
    $I_{NaK}$ & $Na^+ /K^+$ ATPase current    \\ \hline
    $I_{pCa}$ & calcium pump current  \\ \hline 
    $I_{Kb}$ & background rapidly activating potassium current   \\ \hline
    $I_{Nab}$ & background sodium current   \\ \hline
    $I_{Cab}$ & background calcium current    \\ \hline
    
    \end{tabular}

 \hspace{7cm}\caption{\textbf{The various ionic currents in the ORd model; the symbols 
used for these currents follow Ref.~\cite{o2011simulation}. }}
\label{sec:Table3}
\end{center}
\end{table}   

\section{Our Convolutional Neural Network (CNN)}
~\label{sec:CNN}

\begin{figure*}[!ht]
\includegraphics[scale=0.29]{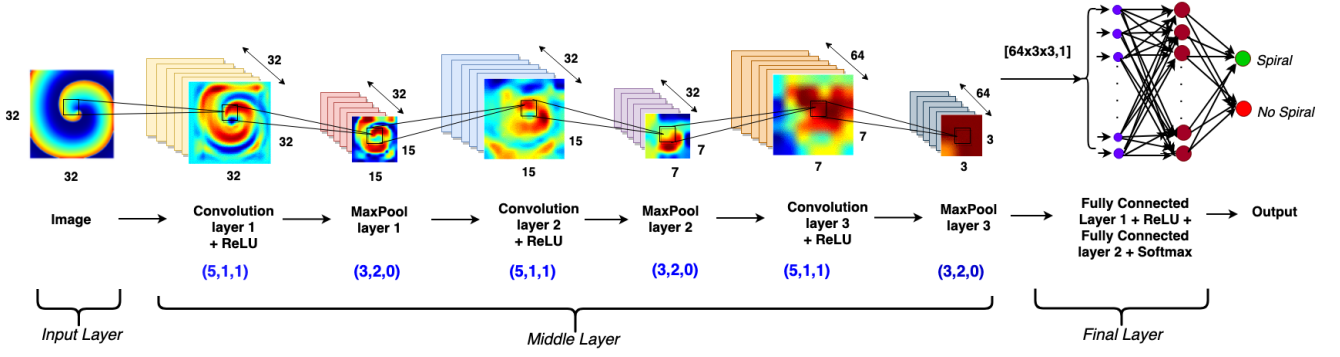}
\caption{Schematic diagram of different layers in our CNN: (a) \textit{Input layer}; 
(b) \textit{Middle layer}; and (c) \textit{Final layer}. The size of the image in 
each layer is controlled by the \textit{filter size} or \textit{pool size} 
$\gamma$, the \textit{stride length} $s$, the padding width $p$, and it is 
determined by the equation $m$ = $\lfloor$($ m'-\gamma+2p $)/$s \rfloor$ +1, where 
$m \times m$ is the image size that is the input from the previous layer. 
Below each layer, we give the values of ($\gamma,\, s, \, p$) in blue. }
\label{fig:cnn}
\end{figure*}

We give below a brief description about the CNN that we develop.  The
pseudocolor image $\tilde{V}$ (see the main paper) is fed into the Convolution
layer 1 through the input layer.  The Convolution layer 1 generates $k$
channels with images that have $m \times n$ pixels from input images of
$m'\times n'$ in $k'$ channels (initially $m' = n' = 32$ and $k' =1$; these
images are enlarged by padding (so that there are $(m' + 2p) \times (n' + 2p)$
pixels) and then combined with filters specified by weights $W_{i,j}^{c,c'}$
and the filter-bias $b^{c}$ as follows:

\begin{eqnarray}
	O^{c,s}_{\mathcal{A}, \ i,j} = \sum_{c'=0}^{k-1} \; \sum_{i',j'=0}^{\gamma-1}
	     \left(I^{c'}_{s\:i+i',s\:j+j'} W^{c,c'}_{i',j'} \right) + b^{c} J_{i,j}, 
\label{eqn:Conv}
\end{eqnarray}

where $s$ is the \textit{stride length}, $c'$ and $c$ are the \textit{channel
numbers} of the input and output, respectively, $ \gamma$ is the \textit{filter
size}, and $m=\lfloor (m'+2 p-1)/s \rfloor + 1$, $n=\lfloor (n'+2 p-1)/s
\rfloor + 1$, $i\in \{0,1,...m-1\}$, $j \in \{0,1,...n-1\}$, where, for $x\in
\mathbb{R}$, $\lfloor x \rfloor$ is the integer part of $x$, and 
$J_{i,j} = 1, \;\: \forall i,j$.

Next, the output from Eq.~(\ref{eqn:Conv}) is fed to ReLU 
that clips the negative values as follows:
\begin{eqnarray}
 O^{c,s}_{\mathcal{B}, \ i,j} = \max\left\{0,O^{c,s}_{\mathcal{A}, \ i,j}\right\}.
	\label{eqn:Relu}
\end{eqnarray}
Then the image from ReLU passes through the MaxPooling layer, which reduces
the image size, but preserves its essential features, via the following operation: 
\begin{eqnarray}
	O^{c,s}_{\mathcal{C}, i,j} = \max_{i',j'=0}^{\gamma'-1} \left\{ O^{c,s'}_{\mathcal{B}, s\:i+i',s\:j
	      +j'} \right\},
\label{eqn:MPool}
\end{eqnarray}
where $i$ and $j$ are defined as in Eq.~(\ref{eqn:Conv}) and $\gamma'$
and $s'$ are, respectively, the \textit{pool size} and the stride length.
The operations in Eqs.~(\ref{eqn:Conv})-(\ref{eqn:MPool}) are repeated 
for each set of Convolution, ReLU, and MaxPooling layers (we have $3$ such sets in 
the Middle layer in Fig.~\ref{fig:cnn}). 

The output from the last MaxPooling layer $O^{c,s}_{\mathcal{C}, \ i,j}$,
which is an $m\times n\times k$ array, is converted into a vector $I_{FL, \ i}$ of
size $mnk \times 1$ and then passed to the fully connected layer (ANNs). We use 
three layers in our ANN, namely: input, hidden, and output layer. From
the ANNs we get:
\begin{eqnarray}
O_{\mathcal{D},\ q} &=& \tanh \left( \sum_{i'=0}^{mnk-1} \left( W_{\mathcal{D}, 
        \ q,i'} \; I_{FL, \ i'} \right) +   b_{\mathcal{D}, \ q} \right) ; \nonumber \\
O_{\mathcal{E},\ q'} &=& \tanh\left(\sum_{q=0}^{\zeta-1} \left( W_{\mathcal{E}, \ q',q} 
        \; O_{\mathcal{D},\ q} \right) +   b_{\mathcal{E}, \ q'}\right); \\
	\tilde{O}_{q} &=& \frac{\exp(O_{\mathcal{E},\ q})}{[\sum_{q'=0}^{\eta-1} 
	     \exp(O_{\mathcal{E},\ q'})]}; 
\label{eq:ANN}
\end{eqnarray}
here $\zeta=256$ is the number of nodes in the hidden layer of the 
ANN, $\eta$ is the number of nodes in the output layer, which corresponds to the
number of classes and $q \in \{0,1,...\eta-1\}$; in our
case $\eta=2$, and the classes are $\mathscr{S}$ \textit{(spiral waves)} and
$\mathscr{NS}$ \textit{(no spiral waves)}.

Initially the weights $W^{c,c'}_{i',j'}$, $W_{FL, \ i,i'}$ , $W_{\mathcal{D}\
q,i'}$ and $W_{\mathcal{E}, \ q',q}$ are chosen to have random values (from a
zero-mean Gaussian distribution with standard deviation $0.01$); and the biases
$ b^{c}$ and $ b_{FL, \ i}$ are taken to be $0$.

\begin{figure}[!ht]
\includegraphics[scale=0.3]{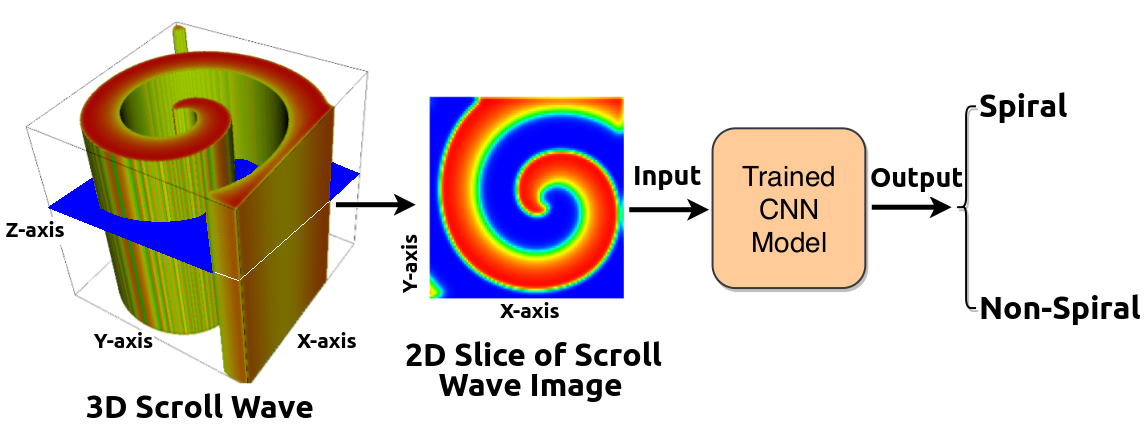}
\caption{An illustration of the detection of a scroll wave by our CNN, which uses a 
two-dimensional (2D) section through the three-dimensional (3D) scroll wave;
.}
\label{fig:3d_slice}
\end{figure}

We have mentioned in the main paper that our CNN can be used to detect scroll
waves in three-dimensional (3D) simualtion domains. An illustration of such
detection by our CNN, which uses a 2D section through the
3D scroll wave is given in Fig.~\ref{fig:3d_slice}.

\section{Control schemes for the elimination of unbroken and broken spiral waves}

We give here brief overviews of three schemes that we use to control unbroken
and broken spiral waves in the mathematical models for cardiac tissue that we
have described above. The first is the mesh-control
scheme~\cite{shajahan2007spiral,shajahan2009spiral,sinha2001defibrillation,pandit2002spiral,nayak2014spiral};
our description here follows that in Ref.~\cite{mulimani2018comparisons}.

\label{sec:defib}

\subsection{Mesh-control scheme}

The development of control schemes for the elimination of unbroken and broken
spiral waves, in mathematical models for cardiac tissue, is of great importance
in searching for low-amplitude defibrillation methods, for the elimination of
life-threatening cardiac arrhythmias, such as ventricular tachycardia (VT) and
ventricular fibrillation (VF). The mathematical analogs for these arrhythmias
are, respectively, unbroken and broken spiral waves.  One such control scheme,
which has been very successful in the elimination of such waves in a variety of
mathematical models for cardiac
tissue~\cite{shajahan2007spiral,shajahan2009spiral,sinha2001defibrillation,pandit2002spiral,nayak2014spiral},
removes spiral waves by the application of currents on a square mesh. In this
scheme, we divide the simulation domain into square cells of dimension
$N_{G}\times N_{G}$ grid points (given a domain of size $N \times N$ grid
points, we must have $N>N_{G}$); we then apply a stimulus current, of amplitude
$50$ pA/pF, for $100$ ms along the edges of the square cells, as in
Ref.~\cite{sinha2001defibrillation}; this eliminates spiral wave, as we show
in top panel of Fig.~\ref{fig:defibr}.

\subsection{Optogenetics-based control scheme:}

\begin{figure}[!ht]
\includegraphics[scale=0.5]{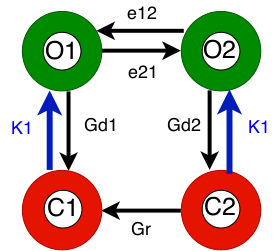}
\caption{A schematic diagram showing the Markov-chain model for {\it
channelorhodopsin}; this consists of two closed states (C1,C2) and two
open states (O1,O2). The rate $\textcolor{blue}{\text{K1}}$ depends on
the strength of the irradiation~\cite{boyle2013comprehensive}.}
\label{fig:markov}
\end{figure}

\begin{figure*}[!ht]
\resizebox{\linewidth}{!}{
\includegraphics[scale=0.1]{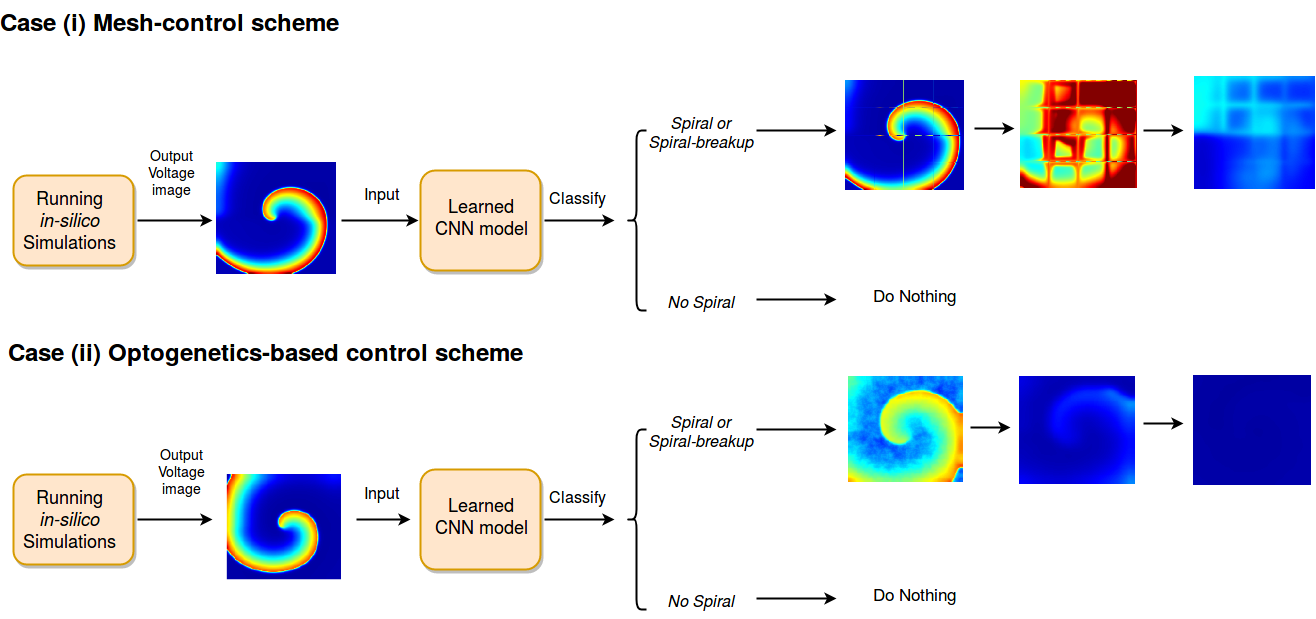}}

\caption{Schematic figures, top and bottom panels, showing the detection of spiral waves 
by our trained CNN model and, therafter, the elimination of these waves
by mesh-control and optogenetics-based-control schemes for the illustrative case of
the LR-I model.}
\label{fig:defibr}
\end{figure*}

Optogenetics is an emerging technique that has been used for controling
excitation waves in cardiac tissue. In particular, it has been used for the
effective elimination of unbroken and broken spiral waves in \textit{in vitro}
and \textit{in silico} studies of cardiac
tissue~\cite{bruegmann2010optogenetic,boyle2013comprehensive,bingen2014light,ambrosi2014cardiac,boyle2015computational,majumder2018optogenetics}.
The light-activated ion-channel protein \textit{channelorhodopsin} is expressed
in a cell (a cardiac myocyte in our case); and the form and intitiation of the
action potential of this cell can be controlled through optical stimulation. In
simulation studies, channelorhodopsin is modelled by a four-state, Markov-chain
model as shown in Fig.\ref{fig:markov}. The equations governing these states,
their rates, and the current throught this channel are given in Eqs. (4)-(9)
(see, e.g., Ref.~\cite{williams2013computational}). In our optogenetics-based
control scheme for the mathematical models of cardiac tissue that we study, we
express the channelorhodopsin in cardiac myocytes, at random positions in the
tissue, such that the percentage of modified cardiac myocytes is $P_f$.  When
our CNN detects an unbroken or broken spiral wave (the class $\mathscr{S}$ in
the main paper) at some time in our numerical simulation, we activate the
channelorhodopsin by irradiating the tissue with light of strength $15$
mW/mm$^2$ for a duration of $\simeq 10-15$ ms.  In bottom panel of
Fig.~\ref{fig:defibr}, we illustrate the elimination of such waves, via this
optogenetics-based method, for the LR-I model; we obtain similar results for
the TP06 and ORd models. The values of $P_f$ and the irradiation strength are
crucial control parameters in the elimination of spiral waves. In our study we
use $P_f = 15\%$, which is the minimum value that leads to successfull control.

\subsection{Gaussian-control scheme applied in LR-I and AP models}

\begin{figure*}[!ht]
\includegraphics[scale=0.5]{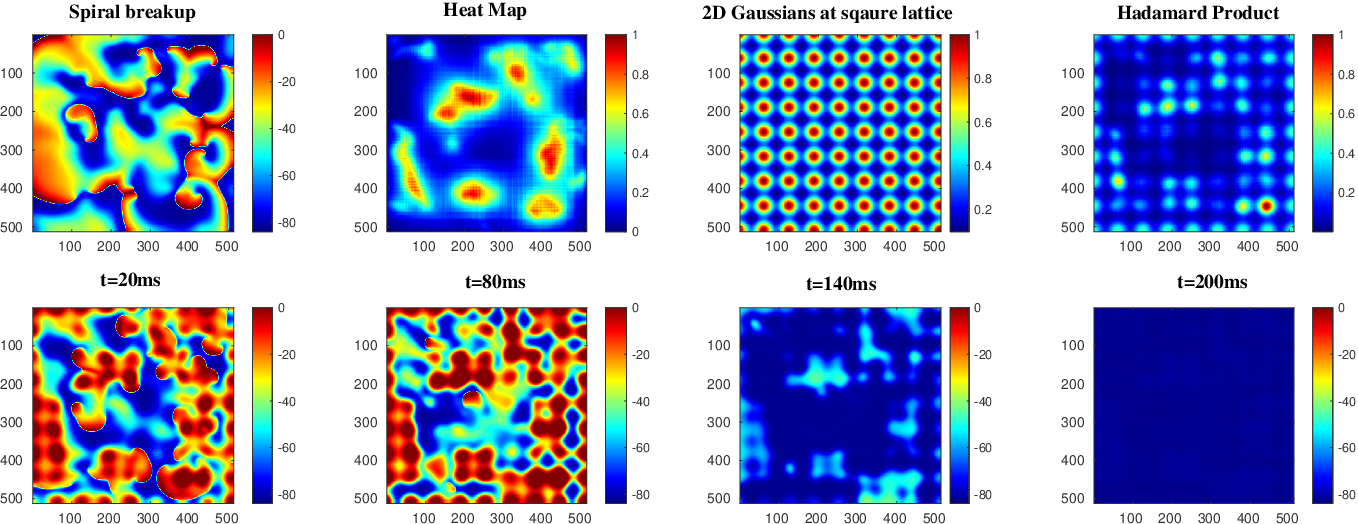}
\put(-256,183) {\bf(C)} \put(-104,183) {\bf(D)}
\put(-352,183) {\bf(B)} \put(-484,183) {\bf(A)}
\put(-226,86) {\bf(G)} \put(-94,86) {\bf(H)}
\put(-342,86) {\bf(F)} \put(-474,86) {\bf(E)}

\caption{Gaussian-control scheme; LR-I model. Pseudocolor plots 
(A) of $V$ (showing broken spiral waves), (B) the corresponding heat map
generated by using our CNN (main paper), (C) 2D Gaussians with $\sigma  =  0.496$
cm, (D) the Hadamard product $\mathcal{P_H}(i,j)$ of 2D Gaussians on a 
square lattice with the heat map (main paper). (E)-(H) Pseudocolor plots 
of $V$, at different instants of time $t$ showing how our Gaussian-control 
scheme (with $I_{def}= 5$ pA/pF and $t_{def} \ = \ 100$ ms) eliminates broken
spiral waves. }

\label{fig:LR}
\end{figure*}

\begin{figure*}[!ht]
\includegraphics[scale=0.465]{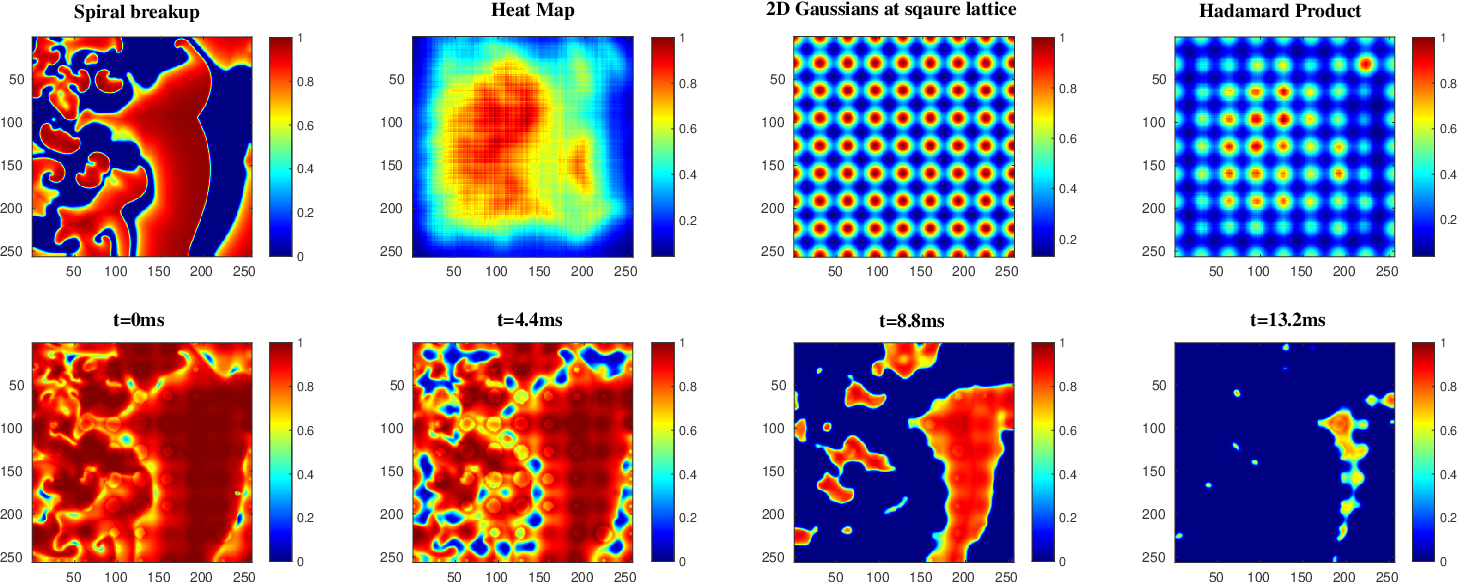}
\put(-256,189) {\bf(C)} \put(-107,189) {\bf(D)}
\put(-348,189) {\bf(B)} \put(-484,189) {\bf(A)}
\put(-226,86) {\bf(G)} \put(-94,86) {\bf(H)}
\put(-342,86) {\bf(F)} \put(-474,86) {\bf(E)}
\caption{Gaussian-control scheme; AP model. Pseudocolor plots 
(A) of $V$ (showing broken spiral waves), (B) the corresponding heat map
generated by using our CNN (main paper), (C) 2D Gaussians with $\sigma  =  0.496$
cm, (D) the Hadamard product $\mathcal{P_H}(i,j)$ of 2D Gaussians on a 
square lattice with the heat map (main paper). (E)-(H) Pseudocolor plots 
of $V$, at different instants of time $t$ showing how our Gaussian-control 
scheme (with $I_{def}= 1.4$ pA/pF and $t_{def} \ = \ 6.6$ ms) eliminates broken
spiral waves.}
\label{fig:AP}
\end{figure*}

In the main paper, we have used the Gaussian-control scheme to eliminate spiral
waves in the TP06 model. We show,  in Fig.~\ref{fig:AP} and Fig.~\ref{fig:LR},
respectively, that this scheme also removes broken spiral waves in the AP
and LR-I models. 

\begin{eqnarray}
I_{ChR2} = g_{ChR2} G(V) (O_1 +\gamma O_2) (V-E_{ChR2}) ;\\
O_1 + O_2 + C_1 + C_2 =1 ;\\
\frac{dC_1}{dt} = G_r C_2 + G_{d1} O_1 - k_1 C_1 ;\\
\frac{dO_1}{dt} = k_1 C_1 - (G_{d1} + e_{12}) O_1 + e_{21} O_2 ;\\ 
\frac{dO_2}{dt} = k_2 C_2 - (G_{d2} + e_{21}) O_2 + e_{12} O_1 ;\\
\frac{dC_2}{dt} = G_{d2} O_2 - (k_2 + G_r) C_2 ;\\
k_1 = \phi_1(F,t) =\epsilon_1 Fp ;\\
k_2 = \phi_2(F,t) =\epsilon_2 Fp ;\\
F= \sigma_{ret} I \frac{\lambda}{(w_{loss} \ {\it h c})} ;\\
\frac{dp}{dt} = \frac{(S_o(\theta) - p)}{\tau_{ChR2}} ;\\
S_0(\theta) = 0.5 (1 + tanh(120(\theta - 0.1))) ;\\
G(V) = \frac{(10.6408 - 14.6408 \  e^{(-\frac{V}{42.7671})})}{V} ;\\
G_{d1} = 0.075 + 0.043 \ tanh(-\frac{(V+20)}{20}); \\
G_r = 4.34 \times 10^5 \ e^{-(0.0211539274 \ V)}. 
\label{eq:chr}
\end{eqnarray}

\bibliographystyle{unsrt}
\bibliography{suppl_reference}